# Interference between Quantum Paths in Coherent Kapitza-Dirac Effect

Nahid Talebi[1] and Christoph Lienau[2]

[1] Stuttgart Center for Electron Microscopy, Max Planck Institute for Solid State Research, 70569 Stuttgart, Germany

[2] Institut für Physik and Center of Interface Science, Carl von Ossietzky Universität Oldenburg, 26111 Oldenburg, Germany

E-mail: n.talebi@fkf.mpg.de

## Abstract

In the Kapitza-Dirac effect, atoms, molecules, or swift electrons are diffracted off a standing wave grating of the light intensity created by two counter-propagating laser fields. In ultrafast electron optics, such a coherent beam splitter offers interesting perspectives for ultrafast beam shaping. Here, we study, both analytically and numerically, the effect of the inclination angle between two laser fields on the diffraction of pulsed, low-energy electron beams. For sufficiently high light intensities, we observe a rich variety of complex diffraction patterns. These do not only reflect interferences between electrons scattered off intensity gratings that are formed by different vector components of the laser field. They may also result, for certain light intensities and electron velocities, from interferences between these ponderomotive scattering and direct light absorption and stimulated emission processes, usually forbidden for far-field light. Our findings may open up perspectives for the coherent manipulation and control of ultrafast electron beams by free-space light.

**Keywords:** Kapiza-Dirac effect, Ponderomotive, Quantum coherent control, Wolkow states, First principles

# Introduction

Coherent control of the shape of quantum wave functions has set the way towards bond-selective chemistry [1], quantum computing [2, 3], and ultrafast control of plasmons [4]. Quantum coherent control originates from the ability to manipulate the interference between quantum paths towards realizing a desired shape of a target wave function by means of shaped, coherent, and/or strong laser excitation. Theoretically, such quantum paths may intuitively be studied using Feynman's path integral approach [5]. This has been instrumental in interpreting observations of above-threshold ionization [6, 7] and high-harmonic generation [8], but also for selective control of the quantum paths using circular and elliptical polarizations [9]. The success of path integrals in understanding coherent control lies in the offered ease of selecting few, physically significant paths from a wealth of mathematically available options. Moreover, important concepts such as the *action* are easily related to classical quantities.

So far, coherent control has mostly been used to shape the wave function of bound-electron states [10], while applications in controlling free electron waves have emerged only recently [11]. Specifically, the inelastic interaction of electrons with optical near-fields can cause attosecond longitudinal electron bunching. Population amplitudes of certain electron states can be controlled precisely by the laser phase in a Ramsey-type experiment [12]. It has been learned that the inelastic processes involved in electron-near-field scattering require moderate laser intensities and are well understood in a minimal coupling Hamiltonian, neglecting ponderomotive forces [13].

For future progress in ultrafast electron microscopy, it appears desirable to avoid the need for matter-based near-field interactions and to simply use light waves in free space to control and shape pulsed electron beams. As a step in this direction, we show here how to use the elastic interaction of free-electron waves with focused, freely-propagating laser-fields to coherently control the transversal distribution of electron wave packets. This is achieved by a generalization of the Kapitza-Dirac (KD) effect [14, 15] to the concomitant utilization of standing-wave and travelling-wave light patterns. In the normal KD effect, electron waves, travelling through a standing–wave pattern of light, are diffracted to transversely populated momentum states at multiples of twice the momentum of free space light. We show that the KD effect can be generalized to the realization of arbitrary momentum states of the electron wavepacket by controlling the interference between quantum pathways originating from distinctly different parts, absorptive and ponderomotive, of the interaction Hamiltonian. This offers fundamentally new degrees of freedom for designing light-controlled phase masks for free-space electron pulses.

# Results

In a normal KD effect, electrons are propagating through a standing–wave pattern of the light intensity and in a direction perpendicular to the momentum of the light. We assume here that the standing wave pattern of light is formed by two counter-propagating light waves with the wave vectors $\vec{k}_1 = +k_{\text{ph}} \hat{y}$ and $\vec{k}_2 = -k_{\text{ph}} \hat{y}$, respectively. Consequently, the electrons are transversely diffracted into distinct orders which are displaced by $\delta k_{\text{el}} = |\vec{k}_2 - \vec{k}_1| = 2k_{\text{ph}}$, where $k_{\text{el}}$ and $k_{\text{ph}}$ are electron and light wavenumbers, respectively (Figures 1a and b). The probability distribution of the diffraction orders has been shown to be dominated by Bessel functions as $J_n(\kappa t)$, where $t$ is the time of propagation of electrons through the standing–wave pattern and the coupling strength $\kappa = e^2 E_0^2 / 8 m_0 \hbar \omega^2$ scales with the square of the electric field amplitude $E_0$ [16, 17]. Here, $e$ and $m_0$ are the elementary charge and mass respectively, $\hbar$ the Planck constant, and $\omega$ the angular frequency of the light wave. Essentially, the spatially varying light intensity introduces a periodic ponderomotive potential along the y-axis which results in a bunching of electron density in regions of low light intensity. After leaving the interaction zone, this electron wave packet transforms into a coherent superposition of plane waves propagating in direction $\vec{k}_{el}^f = \vec{k}_{el} \pm n \delta k_{el} \hat{y}$.

Theoretically, the dynamics of such free-space electron wave packets in KD gratings are commonly described in terms of the Wolkow representation of the electron wavefunction propagating through a vector potential $\vec{A}(\vec{r},t)$. This representation is a special case of the Eikonal approximation where the wavefunction amplitude is taken as constant and spatial gradients of the vector potential are neglected. Then, the wavefunction can be represented as [18]

$$\psi(\vec{r},t) = \psi_0(\vec{r}, t \to 0) \exp\left(-i \frac{e^2}{2\hbar m_0} \int_0^t \vec{A}(\vec{r},\tau)^2 d\tau\right)$$
$$\times \exp\left(-i \frac{e k_{el}}{m_0} \int_0^t A_x(\vec{r},\tau) d\tau\right) \tag{1}$$

where $\psi_0(\vec{r}, t \to 0)$ is the initial state of the electron wave packet, which for a plane wave electron is given by $\psi_0(\vec{r}, t \to 0) = (2\pi)^{-1.5} \exp(i k_{el} x - i \Omega t)$. Here, $\hbar \Omega = 0.5 m_0 v_{el}^2$, is the kinetic energy of the electron, propagating with velocity $v_{el}$ along the x-axis. As usual, the vector potential $\vec{A}$ is related to the electric field, by neglecting the scalar potential, as $\vec{E} = -\dot{\vec{A}}$. Furthermore, by $\vec{A}^2$ we mean $\vec{A} \cdot \vec{A}$. To describe the KD effect, it is sufficient to write the vector potential as $\vec{A} = A_0 \hat{x} \cos(\omega t) \cos(k_{ph} y)$, which by the direct substitution in Eq. (1) can be cast in the form

$$\psi(\vec{r},t) = \frac{1}{(2\pi)^{\frac{3}{2}}} \exp(i k_{el} x - i \Omega t) \exp\left\{-i \frac{e^2 E_0^2}{2 k_{el} \hbar^2 \omega^2} x\right\} \times$$
$$\sum_n i^n J_n\left(\frac{e^2 E_0^2}{2 k_{el} \hbar^2 \omega^2} x\right) \exp(i 2 n k_{ph} y) \tag{2}$$

For this, we have[16] used $E_0 = \omega A_0$ and $k_{ph} = \omega/c$ and have related the delay $\tau$ to the propagation length $x$ as $x = v_{el} t$, with $v_{el}$ being the phase velocity of the electrons [19, 20]. To derive Eq. (2), the temporal oscillation of $A^2 \cos(\omega t)^2$, giving highly oscillatory phase variations at twice the light frequency, has been replaced by its average value $A_0^2/2$. Then, the *ponderomotive phase modulation*, (the first exponential in Eq. (1)) shows a phase variation $\varphi \propto \cos(k_{ph}y)^2 = \frac{1}{2}(1+\cos(2k_{ph}y))$ along the transverse direction. Using the Jacobi-Anger transform [21], this can be expressed as a series of Bessel functions corresponding to the different diffraction orders of the electron beam. This is the last term in Eq. (2). The probability for populating the *n*-th diffraction order then is given as $P_n^{A^2} = J_n^2\left(\frac{e^2 E_0^2}{2m_0\hbar\omega^2}t\right)$. Moreover, the argument of the exponential inside the summation in Eq. (2) can be written as $i2nk_{ph}y = in(\vec{k}_2-\vec{k}_1)\cdot\vec{r}$. This specifies that the final electron wave function can be regarded as a superposition of plane wave electrons occupying the momentum states $|n,-n\rangle$ in the complete basis specified by the momentum states of the two incident light beams, leading to the final momentum of $\hbar\vec{k}_{el}^f = \hbar\vec{k}_{el} + n\hbar\vec{k}_2 - n\hbar\vec{k}_1$.

Importantly, the phase modulation introduced by the second exponential in Eq. (1) is usually discarded in the description of the Kapitza-Dirac effect. As shown in more detail in the Supporting Information, this term causes an ultrafast oscillation of the phase, which for plane wave electrons averages to zero. Hence, the second exponential in Eq. (1) is simply replaced by a multiplicative factor of unity in Eq. (2).

This shows, that in conventional treatments of the Kapitza-Dirac effect, only the ponderomotive forces on the electrons, expressed by the Hamiltonian $H_1 = e^2A^2/2\hbar m_0$, are responsible for the phase modulations and the resulting electron diffraction. In contrast, the absorptive part $H_2 = ek_{el}A_x/m_0$ of the Hamiltonian in Eq. (1) induces only highly oscillatory phase terms, which – in a phase-cycling approach – are neglected. This is justified whenever considering sufficiently long interaction times between electron and field. This interaction can be however shortened dramatically to a few or even less than a single cycle of the light field, by letting electrons interact with spatially confined optical near fields. In this case, it is the absorptive part $H_2$ of the Hamiltion which becomes dominant and causes inelastic electron photon interactions, as for example in photon-induced near-field electron microscopy [13, 16, 19, 22].

The above mentioned approximations lead to the general assumption that free-electrons and light waves cannot inelastically interact in free space. In contrast, for restricted electron-light interaction times, achieved, e.g., by ultrafast laser excitations and/or employing slow-electron pulses, at energies below 100 eV, none of the previous assumptions necessarily holds true. This has recently been evidenced, e.g., by the direct acceleration and bunching of electrons with laser pulses in free space [23, 24]. In this case, $H_1$ may not only contribute to the inelastic but also to the elastic interaction, which will be demonstrated below. As a result, one might observe, even in free space, interferences between quantum paths arising from both parts of the Hamiltonian. We show that, for light gratings formed by optical beams with finite inclination angles, two different paths can reach the momentum recoil of $k_{el,y} = 2nk_{ph}$, provided by the $H_1$ and $H_2$ parts of the Hamiltonian, respectively (see Figs. 1c and d).

These quantum path interferences appear as modulations in the diffraction pattern of the electron waves, different to the diffraction orders observed in the normal KD effect, and hence can - in principle - be observed by a regular position-sensitive detector. To verify these conclusions, we present comprehensive analytical and numerical studies, directly from first principles.

For the sake of simplicity of our analytical model, we first consider the interaction of an electron wave packet with two plane waves of light propagating at angles $\varphi$ and $-\varphi$ with respect to the direction of propagation of the electron (see Figure 1c). In this way, we construct a standing wave pattern along the y-axis and perpendicular to the initial propagation direction of the electron, whereas along the $x$-axis, the superposition of the light beams imposes a traveling wave. The $x$- and $y$-components of the vector potential will be given by $A_x = -2A_0 \sin\varphi \cos(k_{\text{ph}} \sin\varphi\, y) \cos(\omega t + k_{\text{ph}} \cos\varphi\, x)$ and $A_y = -2A_0 \cos\varphi \sin(k_{\text{ph}} \sin\varphi\, y) \sin(\omega t + k_{\text{ph}} \cos\varphi\, x)$, respectively. Inserting them into the Wolkow propagator (Eq. (1)), including only the $H_1$ (pondermotive) Hamiltonian, we obtain

$$\psi(\vec{r},t) = \frac{1}{(2\pi)^{\frac{3}{2}}} \exp(ik_{\text{el}}x - i\Omega t) \exp\left(-i \frac{e^2 E_0^2}{2k_{\text{el}}\hbar^2 \omega^2}\{1 + 2\cos^2(\varphi)\}x\right) \times$$
$$\sum_n i^n J_n\left(\cos(2\varphi)\frac{e^2 E_0^2}{2k_{\text{el}}\hbar^2 \omega^2}x\right) \exp(2ink_{\text{ph}} \sin(\varphi)y) \quad . \tag{3}$$

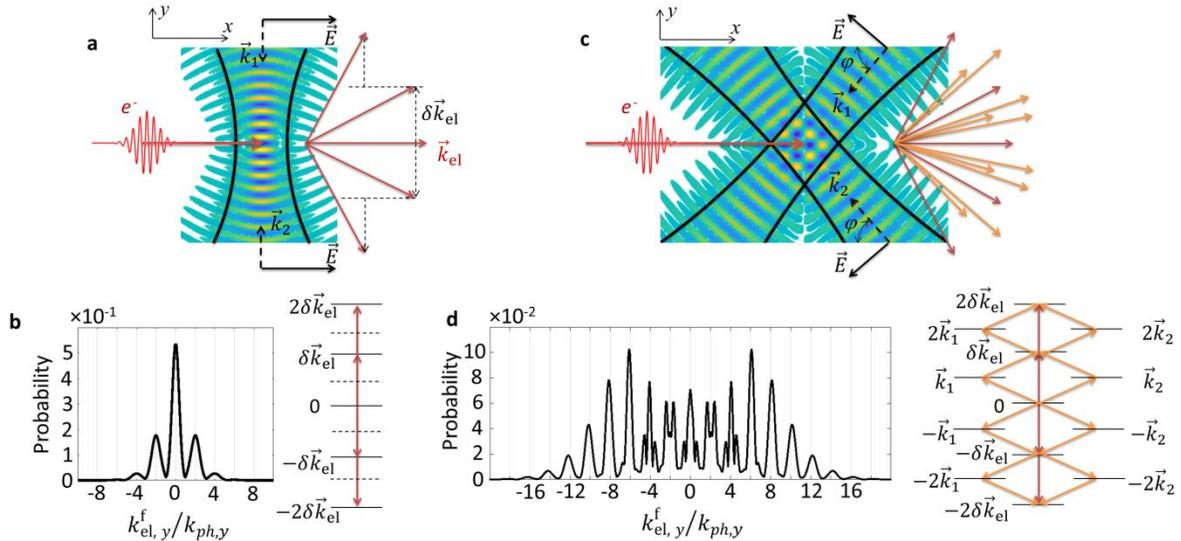

**Figure 1.** Quantum coherent interference paths in the interaction of electron beams with free-space light patterns. (a) When electron beams interact with a standing-wave light pattern in free space, only pondermotive forces are effective, resulting in (b) transverse recoils to the harmonics of $k_y^{el} = n\delta k$. (c) For pulsed electrons synchronized with the laser excitation, interference paths between various parts of the Hamiltonian will result in (d) the formation of an exotic diffraction pattern.

Evidently, for $\varphi = 90°$, Eq. (3) will recast Eq. (2). Interestingly we observe that the probabilities for coupling to the $n$-th diffraction order, with transverse momentum $n\vec{\delta k} = n(\vec{k}_1 - \vec{k}_2)$, now depend sensitively on the inclination angle $\varphi$. Of particular interest is that for $\varphi = 45°$, only the 0-th – diffraction order is populated. In other words, for $\varphi = 45°$, the plane wave electrons will be not diffracted by the light beams. We explain this by the competition between the $H_1^x = e^2 A_x^2 / 2\hbar m_0$ and $H_1^y = e^2 A_y^2 / 2\hbar m_0$ parts of the Hamiltonian, being related to the $\sin^2 \varphi$ and $\cos^2 \varphi$ terms, respectively. Hence, at exactly $\varphi = 45°$, destructive interference between the two paths of the Hamiltonian cancels the diffraction. This cancellation at $\varphi = 45°$, will also occur for incoherent electron beams since their interaction with the pondermotive potential is phase insensitive.

To further investigate the formation of KD diffraction orders under the pure effect of ponderomotive forces, we simulated the interaction of a pulsed electron wave packet with two inclined continuous optical beams with $E_0 = 4 \times 10^{11} V m^{-1}$, $\omega = 62.79 \mathrm{Prads}^{-1}$ ($\lambda_{ph} = 30 \mathrm{nm}$), and $\varphi = 50°$. Such short-wave and intense fields, difficult to realize in the laboratory with present state-of-the-art technology, have been chosen for reducing the numerical complexity of the simulations. A scaling to less intense, longer wavelength pulses will be discussed in more detail below. In these simulations, the interaction Hamiltonian is restricted to $H^{int} = e^2 \vec{A}^2 / 2m_0$; i.e., we neglect the part linearly related to the vector potential. The initial longitudinal and transverse broadenings of the incident electron wave packet are $W_x = 2.4 \mathrm{nm}$ and $W_y = 20 \mathrm{nm}$, respectively, and its initial carrier velocity is $v_{el} = 0.1c$. At this velocity, electron experiences many ponderomotive diffraction orders up to $n = 18$ (see Figure 2a for lower order diffraction probabilities). This electron wave packet, which has a pulse duration of about 0.1 fs, propagates along the *x*-axis, and its overall interaction time with the focused light field is roughly 6 fs, corresponding to about 60 oscillation cycles of the light field (Figure 2b). In the simulations we assume a fixed relative phase between light field and electron wavepacket, as commonly is the case in ultrafast laser-driven electron generation schemes. At the exit of the interaction zone, we notice a gradual formation of symmetric and transversal electron bunching effects with up to 12 distinct lobes (Figure 2c). This generation of KD diffraction orders is better visualized by the momentum representation of the electron wave function (Figure 2d). Longitudinal electron bunching along the direction of the propagation of the electron is not observed, which confirms that inelastic scattering processes are not effective. The generation of up to 28 KD diffraction orders, however, is nicely seen. The number of diffraction orders in the momentum space is significantly larger than the observed number of lobes in the real space. This is related to the fact that many of the observed diffraction orders in the momentum space may result from a pure modulation of the phase of the wave function in the real space - thus they do not become apparent in the amplitude of the wave function in the real space. Considering that the detector is not energy selective, the integration of the probability amplitude along the $k_x$-axis as $P(k_y) = \int dk_x |\psi(k_x, k_y)|^2$ gives the probability distribution of finding the electron in the transversal momentum state $k_y$ (Figure 2e). The peaks of the probability distribution are found at $k_{\text{el},y} = nk_{\text{ph},y} = n\delta\vec{k}_{\text{el}} = nk_{\text{ph}}\sin\varphi$, in agreement with the predictions of our previous Eikonal approximation (Eq. (3)). Noticeable changes in the diffraction spectrum are observed when considering the full interaction Hamiltonian as $H^t = -i\hbar e m_0^{-1}\vec{A}\cdot\vec{\nabla} + 0.5e^2 m_0^{-1}\vec{A}^2$. In the Wolkow approximation, the complete wave function is then formulated as the product of two terms as $\psi(\vec{r},t) = \psi_{\vec{A}^2}(\vec{r},t)\psi_{\vec{A}\cdot\vec{k}}(\vec{r},t)$, where $\psi_{\vec{A}^2}$ is given by Eq. (3) and $\psi_{\vec{A}\cdot\vec{k}}$ is obtained as

$$\psi_{\vec{A}\cdot\vec{k}} = \exp\left(i\frac{ek_{\text{el}}}{m_0}2A_0\sin(\varphi)\cos(k_{\text{ph}}\sin(\varphi)y)\times \int_0^{\delta t}\cos(\omega\tau + k_{\text{ph}}\cos(\varphi)x)d\tau\right). \quad (4)$$

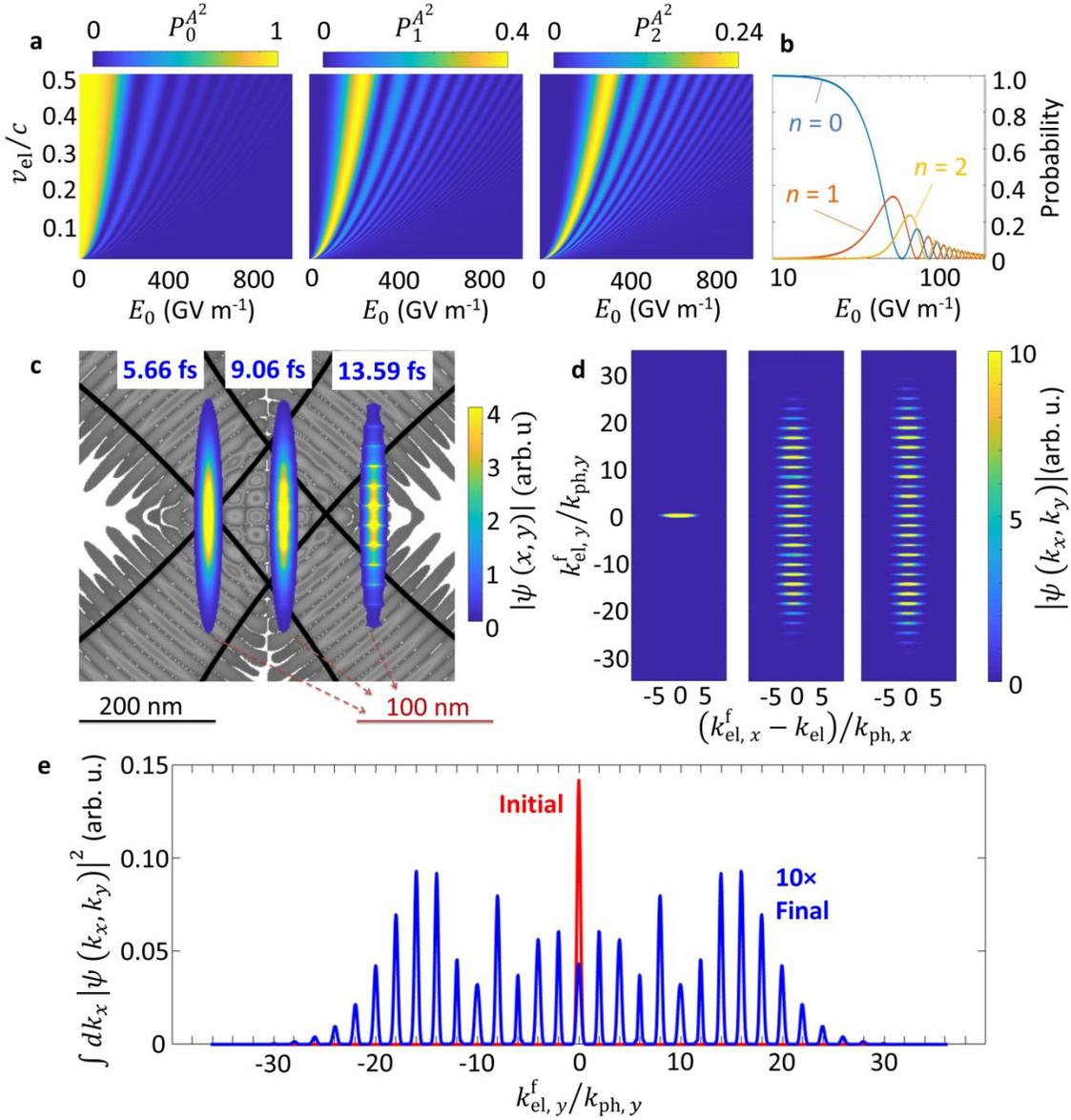

**Figure 2.** Diffraction of electron pulses by the ponderomotive potential of a standing wave light pattern at the wavelength of $\lambda_0 = 30\ nm$. (a) Population densities in the lowest diffraction orders as a function of electron velocity and electric field amplitude of the light beams. (b) Probability of finding the electron in one of the three lowest diffraction orders, calculated for a plane wave electron at a velocity of $v_{el} = 0.03c$. Amplitude of the electron wave function for an electron moving at $v_{el} = 0.1c$ in (c) real space and (d) momentum space, at selected times during the interaction with the standing-wave light field. For clarity, the spatial scale of the electron wave packet in real space is increased by a factor 2 in comparison with that showing electromagnetic field distribution. The electric field amplitude of each optical beam is $E_0 = 400 \mathrm{GVm}^{-1}$ and the wavelength is 30 nm. (e) Transverse momentum distribution of the electron pulse before and after the interaction with the standing-wave field, demonstrating the diffraction of the electron wave into orders $k_{e,y}(n) = n\delta k = n2k_{ph,y}$.

This can be further recast into

$$\psi_{\vec{A}\cdot\vec{k}}(\vec{r},t) = \sum_n\sum_l\sum_m\sum_o \Big\{ i^{-(l+m+n+o)} \times \\ \exp\left(i\{l\vec{k}_1 + o\vec{k}_2\}\cdot\vec{r}\right)\times \\ J_{n-l}(\alpha_s)J_{m-o}(\alpha_s)J_n(\alpha_c)J_m(\alpha_c) \Big\} \quad (5)$$

where $\alpha_c = (ek_{\text{el}}E_0/\omega^2 m_0)\sin(\varphi)(1-\cos(\omega\delta t))$ and $\alpha_s = (ek_{\text{el}}E_0/\omega^2 m_0)\sin(\varphi)\sin(\omega\delta t)$ (see Supplementary Information for details of the derivation). Here, $\delta t = x/v_{\text{el}}$ is the interaction time. More precisely, it is the ratio of the overall interaction time to the temporal period of the laser oscillations, i.e., $\delta t/T$, where $T = 2\pi/\omega$ which cause oscillations in the coupling strengths; i.e. $\alpha_c \propto \sin^2(\pi\delta t/T)$ and $\alpha_s \propto \sin(2\pi\delta t/T)$. This cycling behavior is akin to the Rabi oscillation of the electronic population densities between the photonic states of the momentum ladder set by the free-space electromagnetic excitations [11].

The arguments of the Bessel functions, denoted as $\alpha_{s,c}$ are functions of the parameters of the laser excitation ($E_0$ and $\omega$) as well as the parameters of the electron such as its mass, charge and initial momentum. They scale linearly with the product of initial momentum and field amplitude, $k_{\text{el}}E_0$, in contrast to the arguments obtained for the ponderomotive Hamiltonian, $H_2$, scaling quadratically with the field amplitude. Also, the arguments are well related to the harmonics of the interaction time $\delta t$ relative to the period of the oscillations of the laser excitation, and the angle $\varphi$. Obviously for $\varphi = 0$, $\alpha_c = \alpha_s = 0$ and $\psi_{\vec{A}\cdot\vec{k}} \equiv 1$, regardless of the laser intensity and the interaction time. This is happening because for $\varphi = 0$, there exists no projection of the vector potential over the momentum of the moving electron. Moreover, for interaction times $\delta t = nT$ with $n$ being an integer, no diffraction which can be correlated to the $H_2$ part of the Hamiltonian will occur.

It is of particular interest to notice, that, in contrast to the action of the ponderomotive potential, causing the generation of diffraction orders only at harmonics of $\delta \vec{k} = \vec{k}_1 - \vec{k}_2$, the action of the absorptive part $H_1 = -i\hbar e m_0^{-1} \vec{A} \cdot \vec{\nabla}$ will result in harmonics of the form $l\vec{k}_1 + o\vec{k}_2$, where $l$ and $o$ can now be any arbitrary integer – note that within the Eikonal approximation, $H_1$ can be simplified to $H_1 = \hbar e m_0^{-1} \vec{A} \cdot \vec{k}_{\text{el}}$, as introduced indirectly in the Wolkow solution (Eq. (1)). Thus, this interaction can, in principle, prepare the electron wavepacket in states $|l,o\rangle$ with momentum $\hbar \vec{k}_{\text{el}}^f = \hbar \vec{k}_{\text{el}} + l\hbar \vec{k}_1 + o\hbar \vec{k}_2$. In contrast, only states $|l,-l\rangle$ can be accessed by ponderomotive scattering. The probability of finding the electron after the interaction in state $|l,o\rangle$ can be written as

$$P_{l,o}^{\vec{A} \cdot \vec{k}} = \left| \sum_m \sum_n i^{-(m+n)} J_{n-l}(\alpha_s) J_{m-o}(\alpha_s) J_n(\alpha_c) J_m(\alpha_c) \right|^2. \tag{6}$$

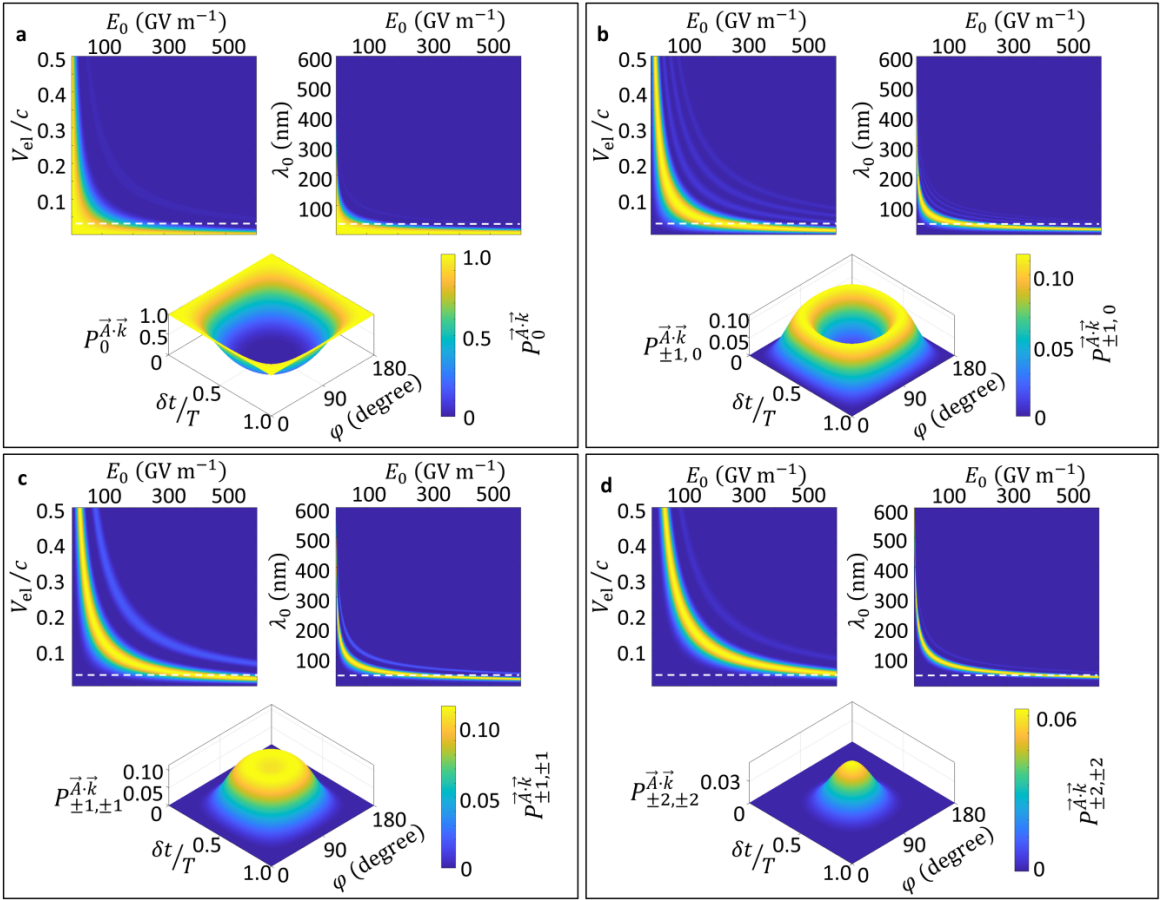

**Figure 3.** Population densities in different diffraction orders resulting from the interaction of electron beams with a propagating light field created by the interference of two inclined light beams under the effect of $\vec{A} \cdot \vec{k}$ Hamiltonian. (a) $|0,0\rangle$, (b) $|0,1\rangle$ and $|1,0\rangle$, (c) $|\pm 1, \mp 1\rangle$, and (d) $|\pm 2, \mp 2\rangle$ final states. Shown in each panel is $P_{l,o}^{\vec{A}\cdot\vec{k}}$ **(top left):** as a function of electron velocity ($v_{el}$) and electric field amplitude ($E_0$) at $\delta t = 0.3T$, $\varphi = 50°$, $\lambda_{ph} = 30\,\text{nm}$, **(top right):** as a function of photon wavelength and electric field amplitude at $\delta t = 0.3T$, $\varphi = 50°$, and $v_{el} = 0.03c$, and **(bottom):** as a function of $\delta t/T$ and $\varphi$, at $\lambda_{ph} = 30\,\text{nm}$ and $v_{el} = 0.03c$.

Interestingly, for fixed values of $\delta t$ and $\varphi$, we notice a different dependence of the population densities $P_{l,o}^{\vec{A} \cdot \vec{k}}$ on electron velocity and electric field amplitude than for $P_n^{\vec{A}^2}$ in the case of purely ponderomotive scattering (compare Figure 3 with Figure 2a). Even at very high field amplitudes, it is still possible to observe the 0-th diffraction order, albeit only for electrons slower than 0.03$c$. The locations of the maxima in general follow a hyperbolic curve in both $(E_0, v_{\text{el}})$ and $(E_0, \lambda_{\text{ph}})$ spaces. This is different for the curves formed under the action of the $H_1$ Hamiltonian, which follow a parabolic shape. Interestingly, for fast electrons, the electron wave function populates higher momentum states already for much lower field amplitude since now the coupling scales linearly with the field amplitude and electron momentum. This behavior is in distinct contrast with the action of the pondermotive potential on the electron, for which larger field amplitudes will be necessary to observe higher order diffractions of the electron by the light. At a specific photon energy and electron velocity, two other parameters, i.e., $\delta t$ and $\varphi$ might be used to control the diffraction orders (Figure 3, bottom panels). For interaction times given by $\delta t = nT$, the Rabi oscillations between the photonic states will cause the population densities to average back to the initial $|0,0\rangle$ state. For this reason, controlling the interaction time appears as a precise way of tuning the population densities. Additionally, the excitation angle $\varphi$ of the photon beams with respect to the direction of the propagation of the electron affects the diffraction orders correlated with both the $H_1$ and the $H_2$ parts of the Hamiltonian, however in a quite different way. Interestingly, at a given interaction time, being set by the longitudinal broadening of the electron wave packet, the angle $\varphi$ can still be used to suppress or release certain diffraction orders.

Importantly, in the presence of large field amplitudes and slow electrons, both parts of the Hamiltonian, i.e. $H_1$ and $H_2$ parts may contribute to the diffraction of the electron beam. This offers a new degree of freedom for manipulating the temporal and spatial structure of the electron pulse by controlling quantum path interferences resulting from the action of both, $H_1$ and $H_2$, on the transition of the electron beam from $|0,0\rangle$ initial state to the final state $|l,o\rangle$. To better understand this, we derive the final electron wave packet using equations (1), (3), and (4) as

$$\psi(\vec{r},t) = \frac{1}{(2\pi)^{\frac{3}{2}}} \exp(-i\Omega t) \times$$

$$\exp\left\{i\left(k_{el} - \frac{e^2 E_0^2}{2\hbar^2 \omega^2 k_{el}} \cos^2(\varphi)\right)x\right\} \times$$

$$\sum_p i^p J_p\left(\frac{e^2 E_0^2}{2\hbar^2 \omega^2 k_{el}} \cos(2\varphi) x\right) \times \qquad . \qquad (7)$$

$$\sum_n \sum_l \sum_m \sum_o \left\{ i^{-(l+n)} i^{-(o+m)} J_n(\alpha_c) J_m(\alpha_c) \times \right.$$

$$\left. J_{n+p-l}(\alpha_s) J_{m-o-p}(\alpha_s) \times \exp\left(i\{l\vec{k}_1 + o\vec{k}_2\}\cdot\vec{r}\right)\right\}$$

As a result the probability of finding the electron at the final $|l,o\rangle$ momentum state will be given as

$$P_{l,o} = \left|\sum_p \sum_n \sum_m i^{p-(l+n+o+m)} J_p\left(\frac{e^2 E_0^2}{2\hbar^2 \omega^2 k_{el,x}} \cos(2\varphi) x\right) J_n(\alpha_c) J_m(\alpha_c) J_{n+p-l}(\alpha_s) J_{m-o-p}(\alpha_s)\right|^2 . \quad (8)$$

This establishes an interesting interference pattern for each given final $|l,o\rangle$ momentum state (Figure 4). One might observe, at specific ranges of field amplitude and electron velocities, quantum coherent interference paths in the diffraction orders. In contrast to previously reported Rabi oscillations in photon-induced near-field electron microscopy [11, 12], the quantum interferences discussed here will occur in free space, and may be utilized as a neat way of controlling the transversal distribution of the electron wave packets, without changing their energy. Interestingly, even in the presence of the $H_2$ interaction, the probability of finding the electrons in the final $|l,-l\rangle$ states is still dominant over $|l,0\rangle$ states. More specifically, the interferences between $H_1$ and $H_2$ interactions are only observed at specific electron velocities. At higher velocities, it is the $H_2$ interaction which is dominating the final diffraction orders of the electron wavepacket, whereas at lower electron velocities, the pondermotive interaction is more pronounced (compare Figure 4 with Figure 2a and Figure 3). This behavior implies that there is a range of electron velocities which allows for the above mentioned interference phenomenon to take place between the $H_1$ and $H_2$ interaction paths. This range of electron velocities strongly depends on applied light frequencies and interaction time as well.

For example, for $\lambda_{ph} = 30\,\text{nm}$, $v_{el} = 0.03c$, and at proper laser field amplitudes, it should be possible to observe an interference between $|0,0\rangle \xrightarrow{H_1} |l,-l\rangle$ and $|0,0\rangle \xrightarrow{H_2} |l,-l\rangle$ paths, and furthermore to coherently

The above-mentioned theory is further benchmarked by directly employing a time-dependent Schrödinger solver based on first principles to analyze the electron-light interaction in the regime where we expect interferences between both terms of the Hamiltonian (see Materials and Methods section for control it by means of the incidence angle $\varphi$ and the laser intensity.

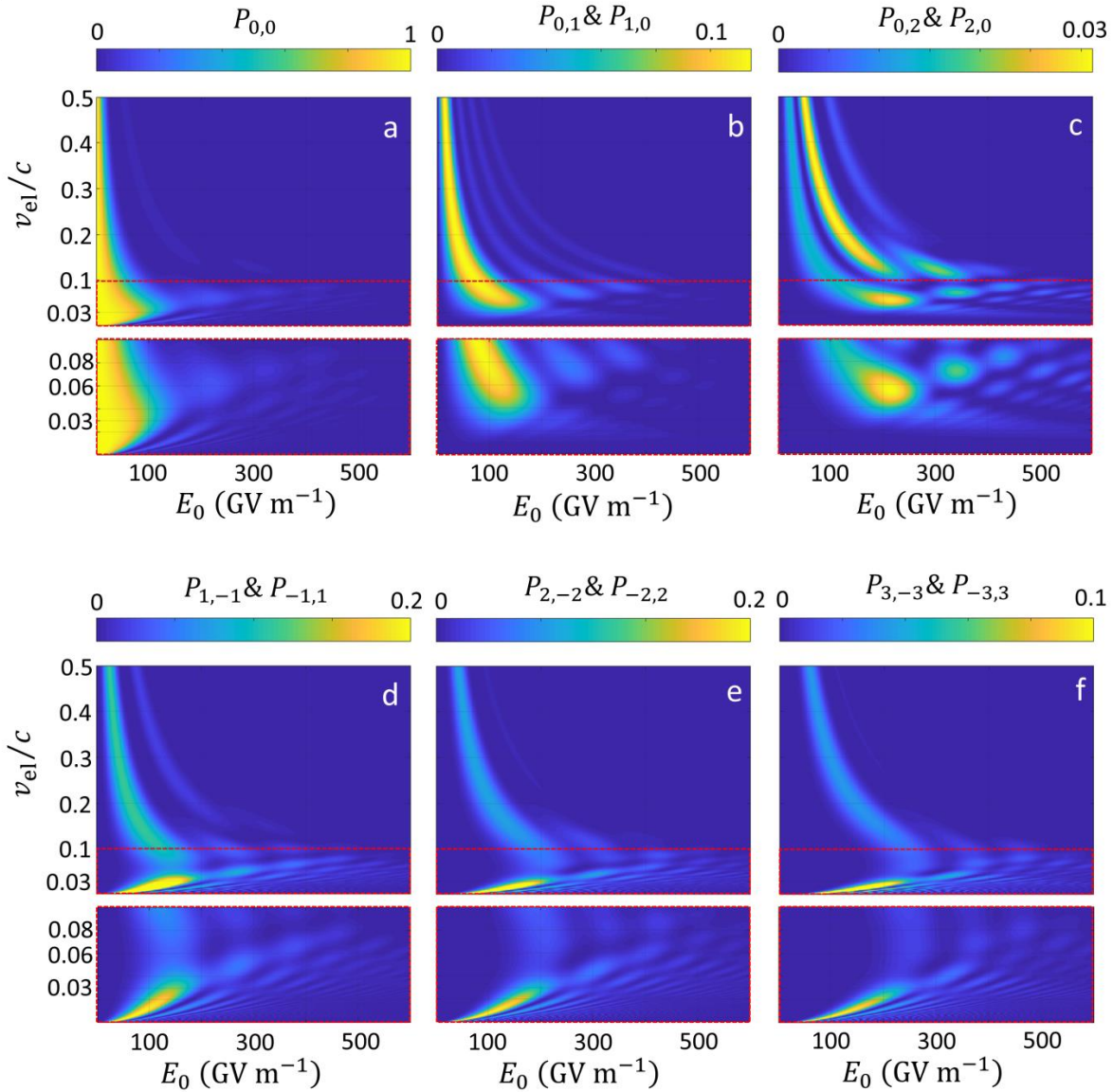

**Figure 4.** Population densities for different diffraction orders (a-d) resulting from the interaction of an electron beam with a propagating light field created by the interference of two inclined Gaussian light beams. The electrons interact with both the ponderomotive $A^2$ and the absorptive $\vec{A}\cdot\vec{k}$ parts of the Hamiltonian. Here $\varphi = 50°$, $\delta t = 0.3T$, and $\lambda_{ph} = 30\,\text{nm}$. Quantum path interferences resulting from the interaction with both terms of the Hamiltonian result in a oscillatory dependence of the electron density on the electric field amplitude.

more details). The electron wave packet has initial longitudinal and transverse broadenings of $W_x = 20\,\text{nm}$ and $W_y = 25\,\text{nm}$, respectively, and its kinetic velocity is $v_{el} = 0.03c$. The laser carrier wavelength and electric-field amplitude are set to $\lambda_{ph} = 30\,\text{nm}$ and $E_0 = 200\,\text{GV}\,\text{m}^{-1}$, respectively and $\varphi = 50°$. Including the complete Hamiltonian, the interaction of the electron wave packet with the laser beams (Figure 5a) is very different from the case of purely pondermotive scattering (Figure 2c). The center of the wave packet is soon depopulated and the wavepacket breaks up into a sequence of sub-peaks along the transverse direction. These subpeaks are becoming increasingly narrower as the wavepacket continues to propagate through the interaction region. The formed bunches of ultrathin wave packets move further away from the interaction region along a direction preassigned by the wave vector of the light beams, hence building up the coherent diffraction orders. In the momentum space, we notice the population of the electron beams prominently into the $|l,-l\rangle$ states, evident by the 5 distinguished bright spots along the transverse direction. However, fainter probability maxima located at $\vec{k}_{el} \pm \vec{k}_1$ and $\vec{k}_{el} \pm \vec{k}_2$ unravel from the non-equilibrium population distribution at $|\pm 1,0\rangle$ and $|0,\pm 1\rangle$ states, as well as higher order $|0,\pm 2\rangle$ and $|\pm 2,0\rangle$ states, which only last for ultrashort times during the interaction (Figure 5b). Moreover, these states provide means to form the desired interference paths between various quantum levels (see Figure 1d), resulting in diffraction orders which are not located at $|l,-l\rangle$ states (Figure 5d and e). These diffraction orders appear as ultrathin electron bunches which are caused by the interference Interestingly, all the diffraction levels will form along the circumference of an Ewald-like sphere (marked in Figure 5d), demonstrating the striking similarity between the elastic interaction of electrons with light and crystalline matter, even though the interaction Hamiltonian is completely different (Figure 5d). The radius of the Ewald sphere is equal to the wavenumber of the electron $k_{el}$, as expected. This further strengthens the fact that the inelastic light-matter interaction, though temporarily happening during the interaction of an electron wave packet with continuous-wave light in vacuum, cannot give rise to a steady-state outcome. This is due to the fact that (i) both light waves specified by wave vectors $\vec{k}_1$ and between the above-mentioned quantum paths (Figure 5c).

$\vec{k}_2$ have exactly similar field strengths, polarizations, and wave numbers, while the only difference between them is the propagation direction, and (ii) the probability amplitude for the transition of the electron wave function from $|0,0\rangle$ momentum state to the 4 final states $|\pm l,\pm l\rangle$, under the effect of $H_2$ interaction and for a given $l$, are all equal. Nevertheless, the transverse (elastic) scattering to the two $|l,-l\rangle$ and $|-l,l\rangle$ states will be caused by both $H_1$ and $H_2$ parts of the Hamiltonian, and the different paths giving rise to the transverse interaction will interfere with each other, causing the appearance of subpeaks in the detected diffraction pattern, which are not displaced by $2nk_{ph}$ (Figure 5 e).

The above mentioned quantum coherent interferences can be controlled by several parameters, including $\varphi$ and $E_0$. Of particular interest is the role of the laser electric-field amplitude. For slow electrons and low laser intensities, the role of the pondermotive force will become dominant. This is to be understood particularly by comparing the Figure 4d with Figure 3 and Figure 2a. As a result one expects to observe diffraction peaks at the order of only $n\delta \vec{k}_{\rm el}$, which is further confirmed by our first-principle calculations (Figure 6). This ponderomotive scattering remains dominant at electric field amplitudes $E_0 \leq 100 \text{GVm}^{-1}$. However for $E_0 > 100 \text{GVm}^{-1}$, the diffraction orders correlated with the $H_2$ part of the Hamiltonian become active as well, and the resulting diffraction pattern is an outcome of the interference between the quantum paths associated with each part of the Hamiltonian. By increasing the field amplitude, the overall number of diffraction orders will be also increased. Interestingly, the highest order diffraction patterns are all at the orders of only $\vec{k}_{el}^{f} = \vec{k}_{el} + 2nk_{ph}\hat{y}$, which demonstrates that the ponderomotive potential still dominates the overall response. Hence the distribution of the diffraction orders due to the $H_2$ part of the Hamiltonian is within a smaller angular range, when compared to the total angular distribution of the diffraction patterns.

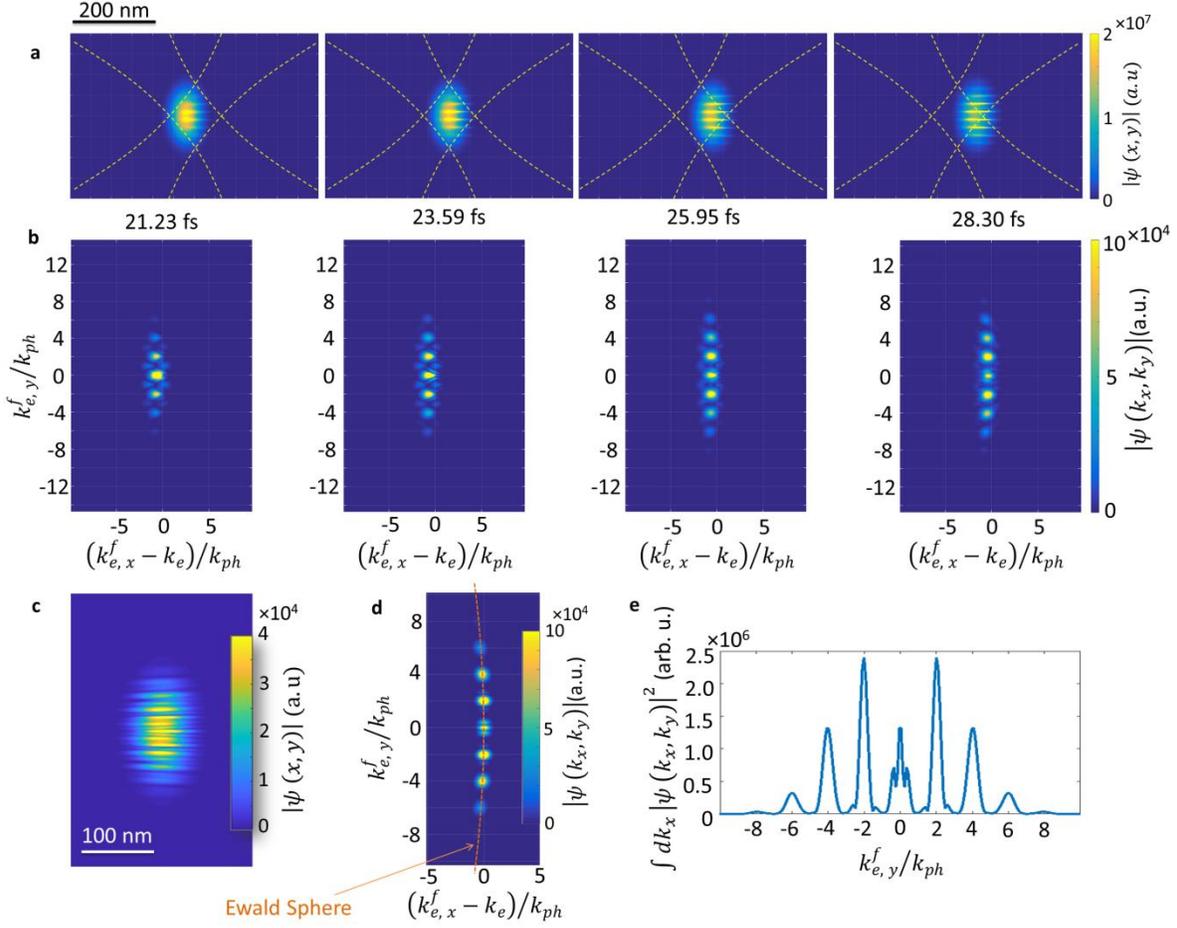

**Figure 5.** Study of the interaction of a pulsed electron beam with two Gaussian optical beams from first-principles. Snapshots of the amplitude of the electron wave function are shown in (a) real space, and (b) momentum space at the interaction time indicated in each frame. The dashed lines are inserted at the locations where the electric field intensity has dropped by a factor of $e^{-1}$. (c) Final shape of the electron wave packet 20 fs after leaving the interaction region, shown by the dashed lines in (a). (d) Final distribution of the electron wave function in momentum space. (e) Probability distribution of the different diffraction orders, as recorded on an electron detector.

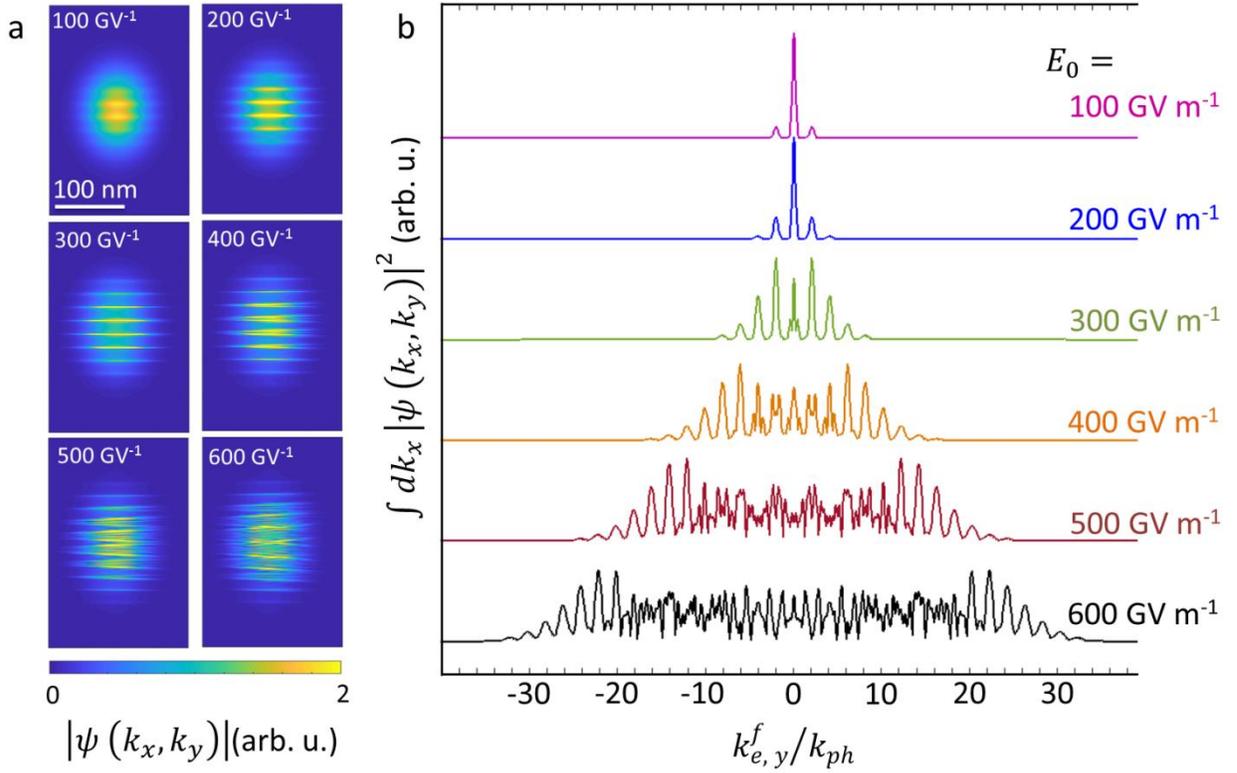

**Figure 6.** Dependence of the diffraction orders on the electric-field amplitude. (a) Spatial distribution of the amplitude of the electron wave function and (b) Integrated probability distributions after the interaction versus the amplitude of the electric-field component of the laser excitation. $\varphi = 50°$, $W_x = 20\,\text{nm}$, $W_y = 25\,\text{nm}$, and $v_{el} = 0.03c$. See the text for description of the parameters.

## Discussions

The short wavelength choice of $\lambda_{ph} = 30\,\text{nm}$ and ultrahigh field amplitudes employed above, were only introduced to meet our simulation resources. These parameters are far beyond what practically can currently be realized in the laboratory. Here, we briefly outline practical approaches towards the realization of quantum path interferences with field strengths and laser wavelengths routinely employed in light–matter interaction experiments. The initial condition for the formation of interferences in quantum paths is to have similar levels of interaction strengths associated with $H_1$ and $H_2$ parts of Hamiltonian. Since $H_1$ is quadratic and $H_2$ is linear in $\vec{A}$, there will be always two values for vector potential amplitudes where $H_1$ and $H_2$ meet each other – one point is $A_0 = 0$ and the other point is related to the electron velocity as $A_0 = 2m_0 v_{el}/e$. Thus, the lower the electron velocity, the lower becomes the required field amplitude (Figure 7a). Additionally, the electric field amplitude is related to the vector potential as $E_0 = \omega A_0$. In other words, it will be possible to employ lower field amplitudes by using oscillating fields at lower frequencies (Figure 7b). For example, by assuming a laser wavelength of $\lambda_{ph} = 1250\,\text{nm}$, the quantum path interferences between $H_1$ and $H_2$ will be observed at electron velocities within $0.01c < v_{el} < 0.1c$ and electric field amplitudes as low as $0.2\,\text{GVm}^{-1} < E_0 < 1.0\,\text{GVm}^{-1}$ (Figure 7c and d). Such conditions can certainly be created by currently available light sources, or by including near-field field-enhancement effects [25].

Dynamics of sideband modulations in the scattering of single-particle electron wavepackets with free-space light can be described as the outcomes of a quantum walk [26] in the discrete momentum states specified by the classical electromagnetic waves. This behavior is akin to the random walks of photons in classical reliable interferometers with low losses and high stability performances [27]. As a result of the coherent action of the unitary operator specified in Eq. (1) on the single electron wavepacket, at each given time the electron will be left entangled between different momentum states. As expected, features of this quantum walk are interferences and boson sampling [27] from a sea of many possible states, as for example selectively selecting few elastic or inelastic scatterings, as a result of such inferences. The ability to dynamically control the outcome of the random walk by few parameters as the polarization, wavelength, intensity, and the inclination angle of the incident Gaussian beams, and particularly by avoiding matter and hence electron-electron and electron-core interactions, make the proposed system a possible promising candidate for bosonic-fermionic random walks and new generation of boson-sampling devices [26, 28, 29].

As a summary, we have generalized the KD effect to the inclusion of two laser fields which propagate at an inclination angle with respect to the electron trajectory. The spatio-temporal behavior of the introduced light waves appears as a standing wave pattern transverse to the electron momentum – similar to the normal KD effect. In addition, a travelling wave pattern forms longitudinal to the initial electron velocity – in contrast with the normal KD effect. We have shown, both theoretically and numerically, that the interaction of the electron wavepacket with the introduced light beams, results in an exotic momentum distribution in the final electron wavepacket, which can be described as the quantum path interferences between two parts of the Hamiltonian, namely ponderomotive and absorptive channels. These interference paths and their coupling strengths can be further controlled in general by the shape of the light waves and in particular by the inclination angles and intensity of the two Gaussian beams, but also by tuning the interaction time and the electron velocity. We anticipate that the aforementioned interference effects can be proposed as new boson sampling device for tailoring the random walk of a single quantum wavepacket in the discrete levels imposed by the momentum of the light.

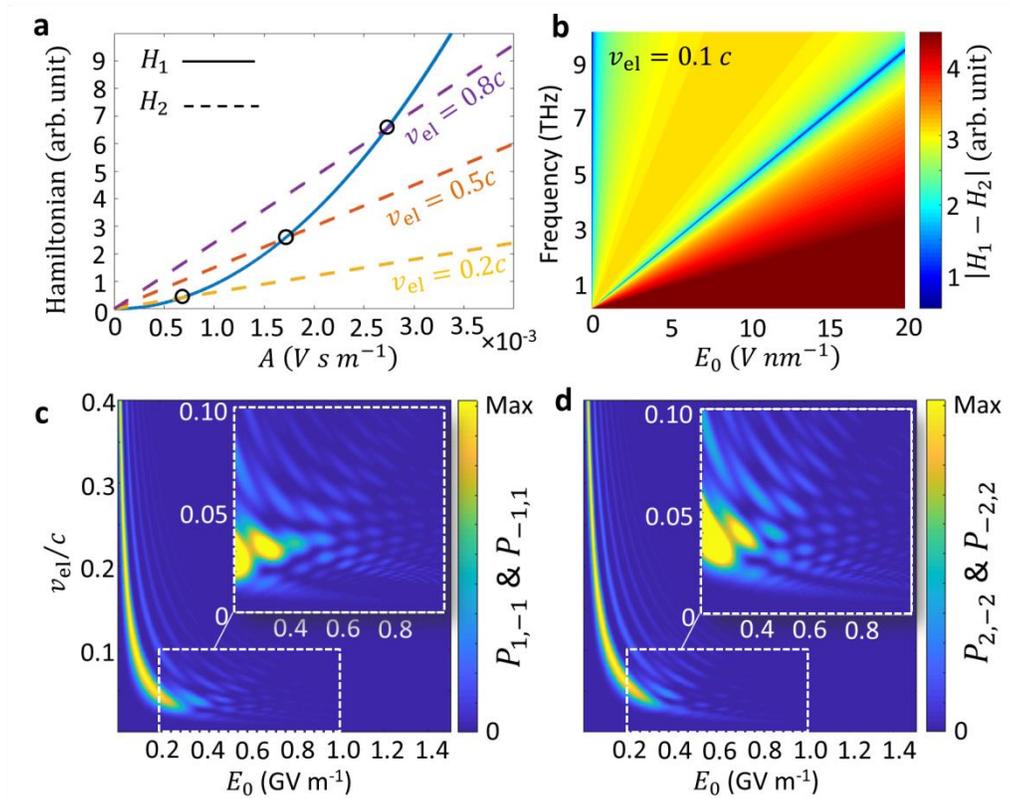

**Figure 7.** Criterion for observation of quantum path interferences between ponderomotive $H_1$ and absorptive $H_2$ parts of the Hamiltonian. (a) Interaction strengths versus the amplitude of the magnetic vector potential, for different electron velocities. (b) $|H_1 - H_2|$ as a function of frequency and electric field amplitude, at the electron velocity of $v_{el} = 0.1c$, Population densities for (c) $|1,-1\rangle$ and $|-1,1\rangle$ and (d) $|2,-2\rangle$ and $|-2,2\rangle$ diffraction orders resulting from the interaction of an electron beam with a propagating light field created by two inclined Gaussian light beams. The electrons interact with both the ponderomotive $A^2$ and the absorptive $\vec{A} \cdot \vec{k}$ parts of the Hamiltonian. Here $\varphi = 50°$, $\delta t = 0.3T$, and $\lambda_{ph} = 1250\,\text{nm}$. Much less electric field amplitudes are required to observe the interference effects as compared with Figure 4.

## Materials and Methods

For our first principle calculations, we have used a time-dependent propagator combined with a pseudospectral Fourier method [30, 31], which conserves the norm of the wave function altogether, and propose a convergent numerical scheme. The accuracy and stability of the numerical method can be controlled by an appropriate choice of the time steps. All the simulations were performed in two-dimensional space. This is particularly rationalized by the choice of polarizations for the fields (magnetic field normal to the simulation domain, electric field lying in the plane). In this way, the classical trajectory of the electron and its experienced quantum mechanical recoils all remain in the simulation plane. Moreover, only a single electron wavepacket is considered. For calculating the optical Gaussian beams, we have used the analytical solutions based on the paraxial approximations, which perfectly model the laser excitations, and is valid for focus regions larger than $1.5\lambda$, where $\lambda$ is the optical wavelength. Here the waist of the introduced optical beams are $2\lambda$. We however have benchmarked this approximation by comparing our results with those obtained using a self-consistent Maxwell-Schrödinger numerical toolbox, and the same diffraction orders and probability amplitudes have been noticed.[32]

## Acknowledgements
NT acknowledges financial support from the European Research Council (ERC Starting Grant NanoBeam). This work was carried out using the high-performance computing facilities of the Carls von Ossietzky Universität Oldenburg.

## Author Information
**Corresponding Author**
E-mail: n.talebi@fkf.mpg.de

**Notes**
The authors declare that they have no conflict of interest.